\documentclass[amsmath,amssymb,%showkeys,
showpacs,twocolumn, nobibnotes,aps,prd]{revtex4-1}

\usepackage{graphicx,amsmath,amssymb}
\usepackage{mathrsfs,eufrak}
\usepackage{dcolumn,color}
\usepackage{bm}

\newcommand{\be}{\begin{equation}}
\newcommand{\ee}{\end{equation}}
\def\ba{\begin{aligned}}
\def\ea{\end{aligned}}

\newcommand{\bea}{\begin{eqnarray}}
\newcommand{\eea}{\end{eqnarray}}

\renewcommand{\hat}[1]{{\widehat #1}}

\begin{document}
%%%%%%%%%%%%%%%%%%%%%%%%%%%%%%%%%%%%%%%%%%%%%%%%%%%%%%%%%%%%%%%%%%%%%%%%%%%%%%%

\title{Vacancy-induced Fano resonances in zigzag phosphorene nanoribbons}
%\title{Theoretical study of vacancy-induced Fano resonances in zigzag phosphorene nanoribbons}
%%%%%%%%%%%%%%%%%%%%%%%%%%%%%%%%%%%%%%%%%%%%%%%%%%%%%%%%%%%%%%%%%%%%%%%%%%%%%%%
\author{M. Amini}\email{msn.amini@sci.ui.ac.ir}
\author{M. Soltani}
\author{M. Sharbafiun}
%%%%%%%%%%%%%%%%%%%%%%%%%%%%%%%%%%%%%%%%%%%%%%%%%%%%%%%%%%%%%%%%%%%%%%%%%%%%%%%
\affiliation{Department of Physics, Faculty of Sciences, University
of Isfahan, Isfahan 81746-73441, Iran}
%%%%%%%%%%%%%%%%%%%%%%%%%%%%%%%%%%%%%%%%%%%%%%%%%%%%%%%%%%%%%%%%%%%%%%%%%%%%%%%

\begin{abstract}
Motivated by recent scanning tunneling microscopy/spectroscopy experiments on probing single vacancies in black phosphorus, 
we present a theory for Fano antiresonances induced by coupling between vacancy states and edge states of zigzag phosphorene nanoribbons~(zPNRs).
To this end, in the first step, using the tight-binding Hamiltonian, we obtain an analytic solution on the lattice for the state associated to a single vacancy located in the bulk phosphorene which shows a highly anisotropic localization in real space.
For a finite zigzag ribbon, in the absence of particle-hole symmetry, 
the localized state induced by vacancies can couple to the wave functions of the edge states
which results in the formation of a new bound state. 
The energy of vacancy bound state lies inside the quasi-flat band composed of edge states when the vacancy locates sufficiently far away from the edge.
Then, we employ the  T-matrix Lippmann-Schwinger approach to obtain an explicit analytical expression for the scattering amplitude
of the edge electrons of a zPNR by the presence of a single vacancy which shows a Fano resonance profile with a tunable dip.
We demonstrate that varying the position of the vacancy produces substantially different effects on the resonance width, resonance energy position, and the asymmetry parameter of Fano line shape.
Furthermore, the validity of the theoretical descriptions is verified numerically by using the Landauer approach.
\end{abstract}
%%%%%%%%%%%%%%%%%%%%%%%%%%%%%%%%%%%%%%%%%%%%%%%%%%%%%%%%%%%%%%%%%%%%%%%%%%%%%%%
%%%%%%%%%%%%%%%%%%%%%%%%%%%%%%%%%%%%%%%%%%%%%%%%%%%%%%%%%%%%%%%%%%%%%%%%%%%%%%%
%%%%%%%%%%%%%%%%%%%%%%%%%%%%%%%%%%%%%%%%%%%%%%%%%%%%%%%%%%%%%%%%%%%%%%%%%%%%%%%
\maketitle
%%%%%%%%%%%%%%%%%%%%%%%%%%%%%%%%%%%%%%%%%%%%%%%%%%%%%%%%%%%%%%%%%%%%%%%%%%%%%%%
%%%%%%%%%%%%%%%%%%%%%%%%%%%%%%%%%%%%%%%%%%%%%%%%%%%%%%%%%%%%%%%%%%%%%%%%%%%%%%%
%%%%%%%%%%%%%%%%%%%%%%%%%%%%%%%%%%%%%%%%%%%%%%%%%%%%%%%%%%%%%%%%%%%%%%%%%%%%%%%
%%%%%%%%%%%%%%%%%%%%%%%%%%%%%%%%%%%%%%%%%%%%%%%%%%%%%%%%%%%%%%%%%%%%%%%%%%%%%%%
%%%%%%%%%%%%%%%%%%%%%%%%%%%%%%%%%%%%%%%%%%%%%%%%%%%%%%%%%%%%%%%%%%%%%%%%%%%%%%%
%%%%%%%%%%%%%%%%%%%%%%%%%%%%%%%%%%%%%%%%%%%%%%%%%%%%%%%%%%%%%%%%%%%%%%%%%%%%%%%
\section{Introduction~\label{Sec01}}
The existence of great transport properties and potential applications of a new semiconducting material, called black phosphorus (BP), has attracted a lot of attention recently~\cite{Li2014,Li2016,Liu2014,Xia2014,Neto-2014,Ling2015}.
BP has a layered structure with van der Waals interaction between individual atomic layers and it is possible to exfoliate a single atomic layer from its bulk which is called phosphorene~\cite{Li2014,Liu2014}.
In this new specimen of two-dimensional~(2D) material, each phosphorus atom is covalently bonded with three nearest neighbors via $sp^3$ hybridization to form a puckered~2D honeycomb structure.
This unique property of anisotropy on one hand, and a tunable and large direct band gap, on the other hand, makes phosphorene a desirable candidate material for different applications with specific electronic, mechanical, thermal, and transport features~\cite{Carvalho2016,Neto2014,Qiao2014,Peeters2014,Guinea2014,Yang2014,Katsnelson2015}.
Furthermore, confining one of the dimensions of phosphorene to a ribbon with the zigzag edge which is called zigzag phosphorene nanoribbon introduces degenerate quasi-flat bands in the middle of the gap detached entirely from the valence and conduction bands~\cite{Ezawa2014,Peeters-skew}. Due to the presence of such edge modes, the quantum transport properties of zPNRs resembles a quasi-one-dimensional~(Q1D) system in this energy window~\cite{Ezawa2014,Asgari2018,Amini2018}.\par
%%%%%%%%%%%%%%%%%%%%%%%%%%%%%%%%%%%%%%%%%%%%%%%%%%%%%%%%%%%%%%%%%%%%%%%%%%%%%%%%%%%%
Despite the vast literature regarding different theoretical and experimental researches on perfect PNRs as
mentioned above, the body of work dedicated to their defective counterparts is surprisingly small.
For instance,  the electronic band structure of defective phosphorene, as well as the formation energy of atomic vacancies, are studied numerically in Ref.~\cite{Hu2015,Wang2015}. 
But among these studies, there is an experiment carried out recently~\cite{Katsnelson2017} which is the observation of atomic vacancies on the surface of bulk BP. 
The spatial distribution of wave function amplitudes around a vacancy which is observed by scanning tunneling microscopy tomographic imaging exhibits strong anisotropy. 
Another significant aspect of the presence of vacancies is their effects on quantum transport through the edges of the zPNRs 
which is computed numerically in Ref.~\cite{Peeters2018} using the tight-binding approach.
It is found that single vacancies can create quasi-localized states which can affect the transport properties of zPNRs by introducing antiresonant dips in the edge conductance of the ribbons. \par
%conductance dips computed numerically in Ref.~\cite{Peeters2018} using the tight-binding approach.  
%It is observed that the vacancies exhibit a
%strongly wave function anisotropy and delocalized charge density.
%Furthermore, the electronic and transport properties of PNRs in presence of atomic vacancies is investigated by means of the tight-binding approach~\cite{Peeters2018}.
%It is found that single vacancies can create quasilocalized states which can affect the transport properties of zPNRs by introducing
%antiresonant dips in the edge conductance of the ribbons. \par
%%%%%%%%%%%%%%%%%%%%%%%%%%%%%%%%%%%%%%%%%%%%%%%%%%%%%%%%%%%%%%%%%%%%%%%%%%%%%%%%%%%%%%
However, the results outlined in the previous paragraph bring up an important question of whether the conductance dips computed for the electronic transmission through the edges of zPNR is due to the coupling of anisotropic vacancy state to the continuum states of the edges resulting in the so-called Fano antiresonance in the wave propagation. 
%Another significant aspect in the study of quantum transport is the Fano resonances induced by defects.
Indeed, Fano resonance~\cite{Fano} is one of the most important resonance mechanisms in which a discrete level couples to an energetically close-by continuum of states giving rise to new scattering paths.  
%Given, the effects of atomic vacancies on the electronic and transport properties of zPNRs  outlined in the previous paragraph, 
It is quite predictable that a single vacancy can introduce a new localized state within the energy window of the edge modes.
Therefore, a possible scenario to understand the corresponding conductance dips of vacant zPNRs can be due to the coupling of the edge states forming a quasi-flat band to the vacancy wave function.
In this work, we present the theory of such resonance scattering due to the vacancy defect leading to Fano resonance in the transmission of edge channels in zPNRs. 
To this end, we first use a tight-binding model to obtain and analyze in detail the localized state induced by a vacancy defect  
in the energy interval of the edge band in zPNRs.
Then, we employ the general scattering theory based on the Green's function approach and the
Lippmann-Schwinger equation to investigate the antiresonance dips in the electronic transmission through the edge states of zPNRs due to the presence of such a localized state.
%%%%%%%%%%%%%%%%%%%%%%%%%%%%%%%%%%%%%%%%%%%%%%%%%%%%%%%%%%%%%%%%%%%%%%%%%%%%%%%%%%%%%%

The paper is organized as follows. In section~\ref{Sec02}, we provide the rigorous derivation of localized wave function around a single vacancy in phosphorene using the tight-binding Hamiltonian.
In this section, the energy shift of such a bound state  due to the absence of electron-hole symmetry is explained. 
In section~\ref{Sec03}, we generally introduce the
theory of a Fano antiresonance induced by the coupling of an impurity state to a continuum of states using the T-matrix Lippmann-Schwinger approach. 
We employ this formalism in section~\ref{Sec04} to obtain the Fano antiresonance dips in the electronic transmission through the edge states of zPNR which is induced by a single vacancy near the edge.
Section~\ref{Sec05} is devoted to discussion on the obtained
results and, finally, we wrap up the paper with the summary and concluding remarks in section~\ref{Sec06}.

%%%%%%%%%%%%%%%%%%%%%%%%%%%%%%%%%%%%%%%%%%%%%%%%%%%%%%%%%%%%%%%%%%%%%%%%%%%%%%%
%%%%%%%%%%%%%%%%%%%%%%%%%%%%%%%%%%%%%%%%%%%%%%%%%%%%%%%%%%%%%%%%%%%%%%%%%%%%%%%
%%%%%%%%%%%%%%%%%%%%%%%%%%%%%%%%%%%%%%%%%%%%%%%%%%%%%%%%%%%%%%%%%%%%%%%%%%%%%%%
%%%%%%%%%%%%%%%%%%%%%%%%%%%%%%%%%%%%%%%%%%%%%%%%%%%%%%%%%%%%%%%%%%%%%%%%%%%%%%%
%%%%%%%%%%%%%%%%%%%%%%%%%%%%%%%%%%%%%%%%%%%%%%%%%%%%%%%%%%%%%%%%%%%%%%%%%%%%%%%
%%%%%%%%%%%%%%%%%%%%%%%%%%%%%%%%%%%%%%%%%%%%%%%%%%%%%%%%%%%%%%%%%%%%%%%%%%%%%%%%
\section{Localized states around vacancies in bulk phosphorene\label{Sec02}}

\begin{figure}[t]
\centering
\includegraphics[scale=0.8]{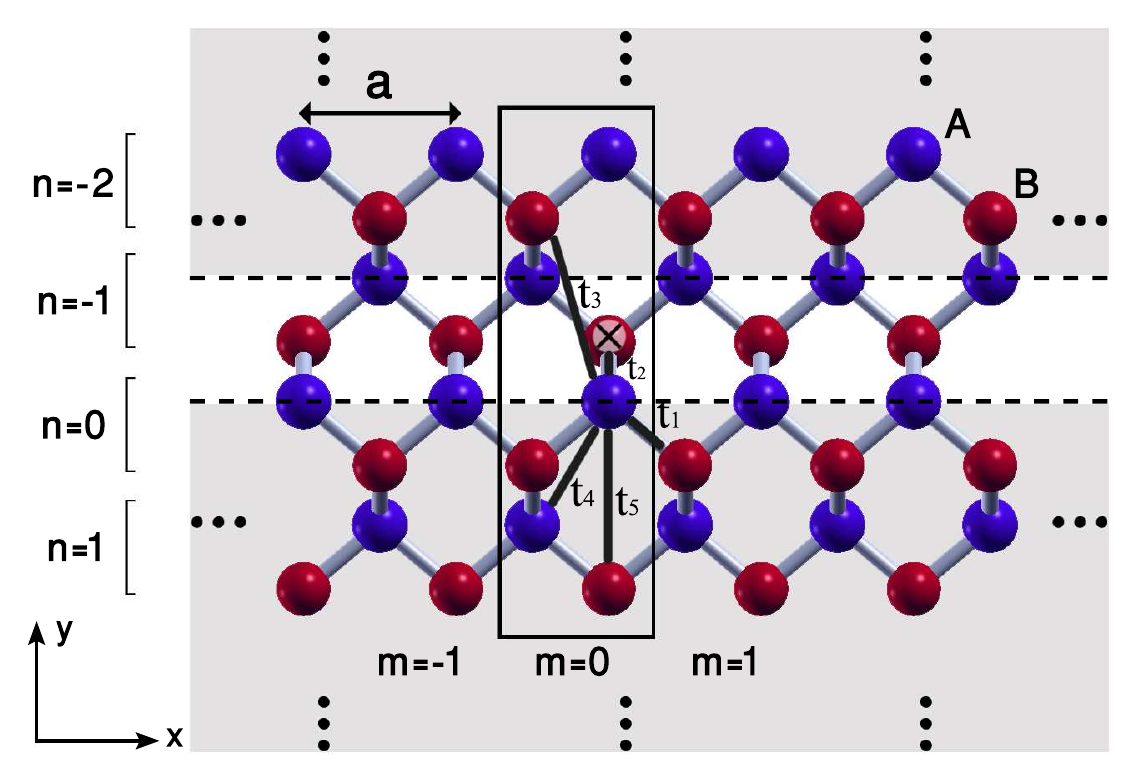}
\caption{(Color online). Schematic view of the lattice structure of phosphorene in real space.
The lattice constant is $a$, $m$ is the index of the vertical armchair chain along the horizontal direction and
$n$ labels the horizontal zigzag chain along the vertical direction.
This lattice consists of two different types of sublattices denoted by $A$ and $B$ and
a single vacancy which is located on site $(m=0,n=-1,B)$ and denoted by $\times$ symbol.} \label{Fig01}
\end{figure}

Phosphorene consists of a non-planar puckered honeycomb lattice of phosphorus atoms with two in-equivalent sublattices denoted by $A$ and $B$ as shown in Fig.~\ref{Fig01}.
We choose the $x$ and $y$ axes along the zigzag and armchair chains respectively. 
Now, it is possible to describe each lattice site by $(m,n,\nu)$ where $m$ and $n$ are armchair and zigzag chain numbers respectively and $\nu = A, B$.
The energy band structure of phosphorene can be well described using a tight-binding model~\cite{Ezawa2014,Rudenko2014}
\begin{equation}
{\hat H} =  \sum_{\langle i,j \rangle} t_{ij} c_{i}^\dagger c_j + h.c. , \label{EQ1}
\end{equation}
where $\langle i,j \rangle$ represents summation over only considered neighbors.
Here, $t_{ij}$ is the hopping energy between sites $i$ and $j$, $c^\dagger_i(c_i)$ is the electron creation (annihilation)  operator in site $i$, and $h.c.$ stands for hermitian conjugate.
It is shown that considering only five hopping integrals $t_1 =-1.220$~eV, $t_2 = 3.665$~eV, $t_3 =-0.205$~eV, $t_4 =-0.105$~eV, and $t_5 =-0.055$~eV in this model, which is shown in Fig.~\ref{Fig01},  provides a reasonable description of the phosphorene band structure~\cite{Rudenko2014}.
However, for analytical calculations, it is reasonable to
neglect hopping terms $t_3$ and $t_5$ which have a smaller contribution to the Hamiltonian in comparison to $t_1$ and $t_2$ terms.
It is important to note that the presence of $t_4$ breaks both electron-hole and lattice inversion symmetries of phosphorene and we will take into account its effect perturbatively later. Therefore, we can rewrite the above Hamiltonian keeping only the important terms of first, second and fourth nearest neighbors as
\begin{equation}
\label{EQ2}
\begin{split}
{\hat H} &  = {\hat H_0}+{\hat H_1}, \\
{\hat H_0} &  = t_1 \sum_{\langle i,j \rangle_\text{1st}}  c_{i}^\dagger c_j + t_2 \sum_{\langle i,j \rangle_\text{2nd}}  c_{i}^\dagger c_j + h.c., \\
{\hat H_1}  &  = t_4 \sum_{\langle i,j \rangle_\text{4th}}  c_{i}^\dagger c_j + h.c.
\end{split}
\end{equation}

%%%%%%%%%%%%%%%%%%%%%%%%%%%%%%%%%%%%%%%%%%%%%%%%%%%%%%%%%%%%%%%%%%%%%%%%%%%%%%%
%%%%%%%%%%%%%%%%%%%%%%%%%%%%%%%%%%%%%%%%%%%%%%%%%%%%%%%%%%%%%%%%%%%%%%%%%%%%%%
%In the absence of vacancies, the Hamiltonian can be simplified using the translational symmetry along the vertical direction. The states can be clas-
%sified by the momentum (in units of 1=a, with a the lattice
%constant) along the vertical axis

Without loss of any generality, let us now consider a single vacancy placed at  site $(0,-1,B)$. This is equivalent to turning the hopping terms that connect this site to its neighbors off.
In the presence of particle-hole symmetry, $t_4=0$, this implies the appearance of a new zero energy state which has only non-zero amplitudes on the $A$ sublattice~\cite{Pereira2008} and we call it $|\Psi_v^{A}\rangle$.
Therefore, it should satisfy the condition 
\be
\hat{H_0}|\Psi_v^{A}\rangle =0.
\label{EQ3}
\ee
Later, we will calculate the effect of $t_4$ perturbatively which shifts this energy toward the negative values.

Indeed, a single vacancy can divide the phosphorene plane into an upper semi-infinite ribbon with beard edge and a lower one with the zigzag edge which is shown by black dashed lines in Fig.~\ref{Fig01}.  
The upper semi-infinite ribbon with beard edge does not support localized edge states~\cite{Ezawa2014}. So, the vacancy wave function amplitudes are zero in the upper half-plane.
On the other hand, due to the absence of hopping between site $(0,0,A)$ and the vacant site, a  possible solution in the lower half-plane can start with non-zero amplitude on this site, namely $\langle 0,0,A|\Psi_v^{A}\rangle\neq0$.
Also Eq.~(\ref{EQ3}) implies that $t_2\langle 1,1,A|\Psi_v^{A}\rangle+ t_1\langle 0,0,A|\Psi_v^{A}\rangle =0$ which results in $\langle 1,1,A|\Psi_v^{A}\rangle =(-t_2/t_1) \langle 0,0,A|\Psi_v^{A}\rangle$. With the same kind of reasoning it is evident that 
$\langle 0,1,A|\Psi_v^{A}\rangle =(-t_2/t_1) \langle 0,0,A|\Psi_v^{A}\rangle$.
Next, one can apply the same procedure to obtain the amplitudes of the wave function on the next zigzag chains which we do not pursue it here.
%, but we do not present them here.
Instead, we focus on a more systematic way of constructing the vacancy wave function
~\cite{Pereira2006,Vozmediano} 
%~\cite{Pereira2006,Vozmediano,Dutreix2013} 
below.

\begin{figure}[t]
\centering
\includegraphics[scale=0.7]{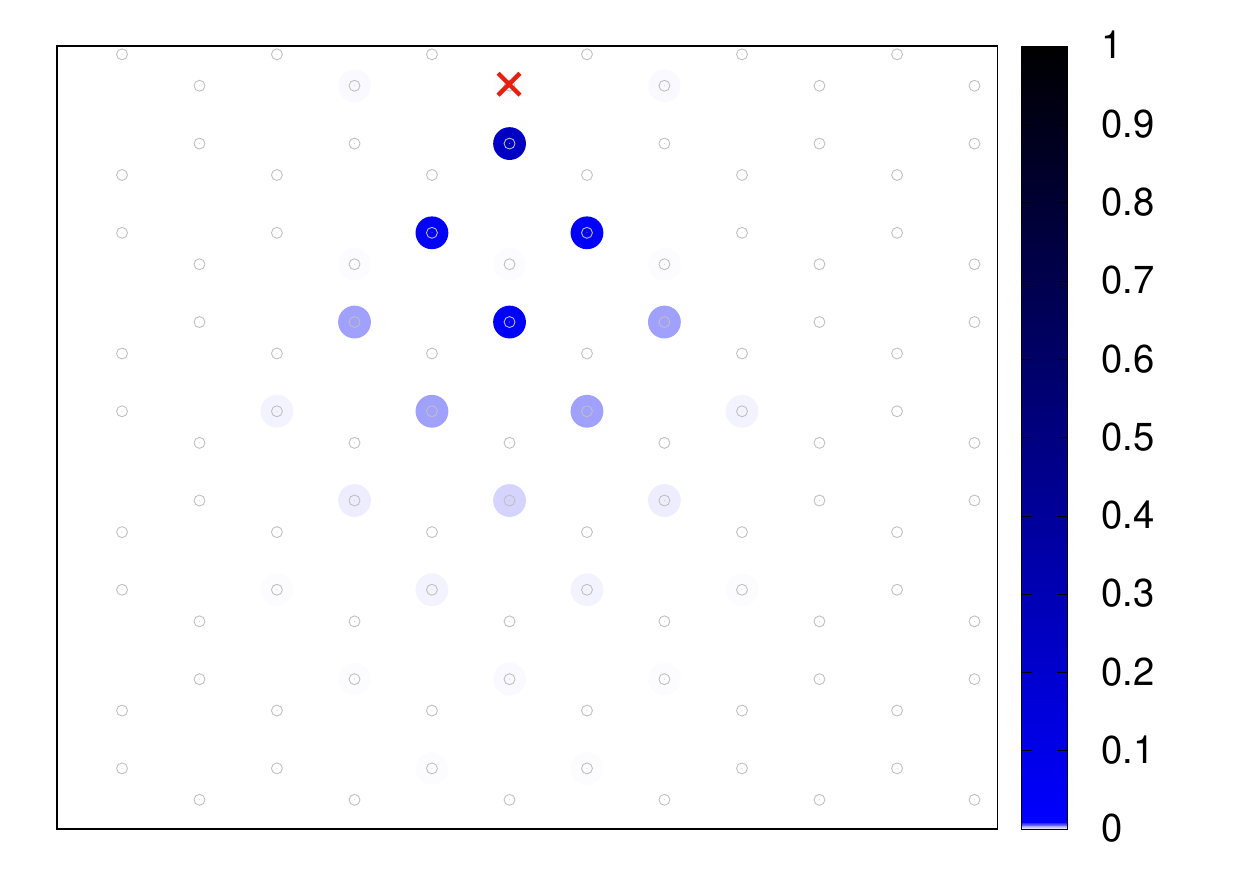}
\caption{(Color online). 
The wave function components of an impurity state around a single vacancy which is located on the site denoted by $\times$ symbol in bulk phosphorene.} \label{Fig02}
\end{figure}

%As we just discussed, the vacancy wave function has only non-zero amplitudes in the lower half-plane.
Let us consider the lower semi-infinite zigzag ribbon in which the amplitudes of the vacancy wave function is non-zero.
The typical edge states of this ribbon have the following wave functions~\cite{Amini2018}
%Furthermore, the lower semi-infinite zigzag ribbon, supports typical edge states with zero energy (when $t_3=t_5=0$) which is given as~\cite{Amini2018}
\be
|\Psi^A(k)\rangle=\frac{1}{\sqrt{\pi}}\sum_{m,n} \alpha^n(k) \gamma(k)e^{ik(m-\delta_n)}|m,n,A\rangle,
\label{EQ4}
\ee
where $k$ is the wave vector measured in units of $1/a$ (and $a$ is the lattice constant) along the horizontal axis.
Here, $\delta_n$ is a constant value equal to 0~(0.5) for even~(odd) $n$, $\alpha(k)=-2(t_1/t_2)\cos{(k/2)}$, and
$\gamma(k)=\sqrt{1-\alpha^2(k)}$. Also, the superscript $A$ denotes the type of sublattice in which the wave function is localized and $n$ is the zigzag chain number measured from the edge (starting from zero).
Therefore, the idea is to express the zero energy vacancy state as a superposition of these edge states with zero components everywhere along the zigzag edge except on site $(0,0,A)$. The following combination of the edge states can satisfy such a condition
\begin{equation}
\label{EQ5}
|\Psi_v^{A}\rangle = c \int_{-\pi}^{\pi} \gamma^{-1}({k})|\Psi^A(k) \rangle dk,
\end{equation} 
in which $c$ is the normalization coefficient given by
\begin{equation}
\label{EQ6}
c^{-2} = \int_{-\pi}^{\pi} \gamma^{-2}(k) dk.
\end{equation} 
Now, we can obtain the amplitudes of the wave function in Eq.~(\ref{EQ5}) on different lattice sites. 
Let us start from 
\be
\label{EQ7}
\begin{split}
\langle m,0,A|\Psi_v^{A}\rangle &= c \int_{-\pi}^{\pi} \gamma^{-1}(k) \langle m,0,A|\Psi^A(k) \rangle  dk \\ 
&= c \int_{-\pi}^{\pi} e^{ikm} dk = c \delta_{m,0}.
\end{split}
\ee
This ensures our desired form of the wave function on the $n=0$ zigzag chain.
Clearly, we can go on and calculate the wave function amplitudes on the $n=1$ zigzag chain as
\be
\label{EQ8}
\begin{split}
\langle m,1,A|\Psi_v^{A}\rangle &= c \int_{-\pi}^{\pi} \gamma^{-1}(k) \langle m,1,A|\Psi^A(k) \rangle  dk \\ 
&= c \int_{-\pi}^{\pi} dk \cos{(k/2)}\left[e^{ik(m-1/2)}+e^{-ik(m+1/2)}\right] \\
&= (-t_2/t_1)c \left[\delta_{m,0}+\delta_{m-1,0}\right].
\end{split}
\ee
This is equivalent to the results of the first method which we described before. 
%but we do not present here the calculations and we only represent the resulting wave function amplitudes in Fig.~\ref{Fig02}.
Fig.~\ref{Fig02} represents the corresponding wave function amplitudes around a vacancy located on the site which is denoted by $\times$ symbol. 
As it is shown, the vacancy wave function is highly anisotropic and non-zero in a region like a triangle over the apposite sublattice in which the vacancy is located.
This vacancy wave function resembles the topological boundary state emerges at the corners of a nanodisk~\cite{Ezawa2018}.  
%is ontained by numerical computations of Ref.~\cite{Amini2018}.  

%%%%%%%%%%%%%%%%%%%%%%%%%%%%%%%%%%%%%%%%%%%%%%%%%%%%%%%%%%%%%%%%%%%%%%%%%%%%%%%
%%%%%%%%%%%%%%%%%%%%%%%%%%%%%%%%%%%%%%%%%%%%%%%%%%%%%%%%%%%%%%%%%%%%%%%%%%%%%%

Now, It is straightforward to calculate the energy correction to this zero-energy mode by the presence of $\hat{H_1}$ term in Eq.~(\ref{EQ2}). 
In doing so, we employ the first order perturbation theory which leads to
\begin{equation}
\label{EQ9}
E_{00} = \langle \Psi_v^{A}|\hat{H_1}|\Psi_v^{A}\rangle=c^2\int_{-\pi}^{\pi} \gamma^{-2}(k) 2 t^{\prime} \cos^{2}{(\frac{k}{2})} dk
\end{equation} 
with $t^{\prime}=(2t_4t_1)/t_2$ in which the integration can be carried out using contour integration~\cite{Amini2018}.
Eqs.~(\ref{EQ9}) and (\ref{EQ5}) will be used in the next section to calculate the electronic transmission through the edge states of a zPNR in presence of a single vacancy. 

%%%%%%%%%%%%%%%%%%%%%%%%%%%%%%%%%%%%%%%%%%%%%%%%%%%%%%%%%%%%%%%%%%%%%%%%%%%%%%%
\begin{figure}[b]
\centering
\includegraphics[scale=0.7]{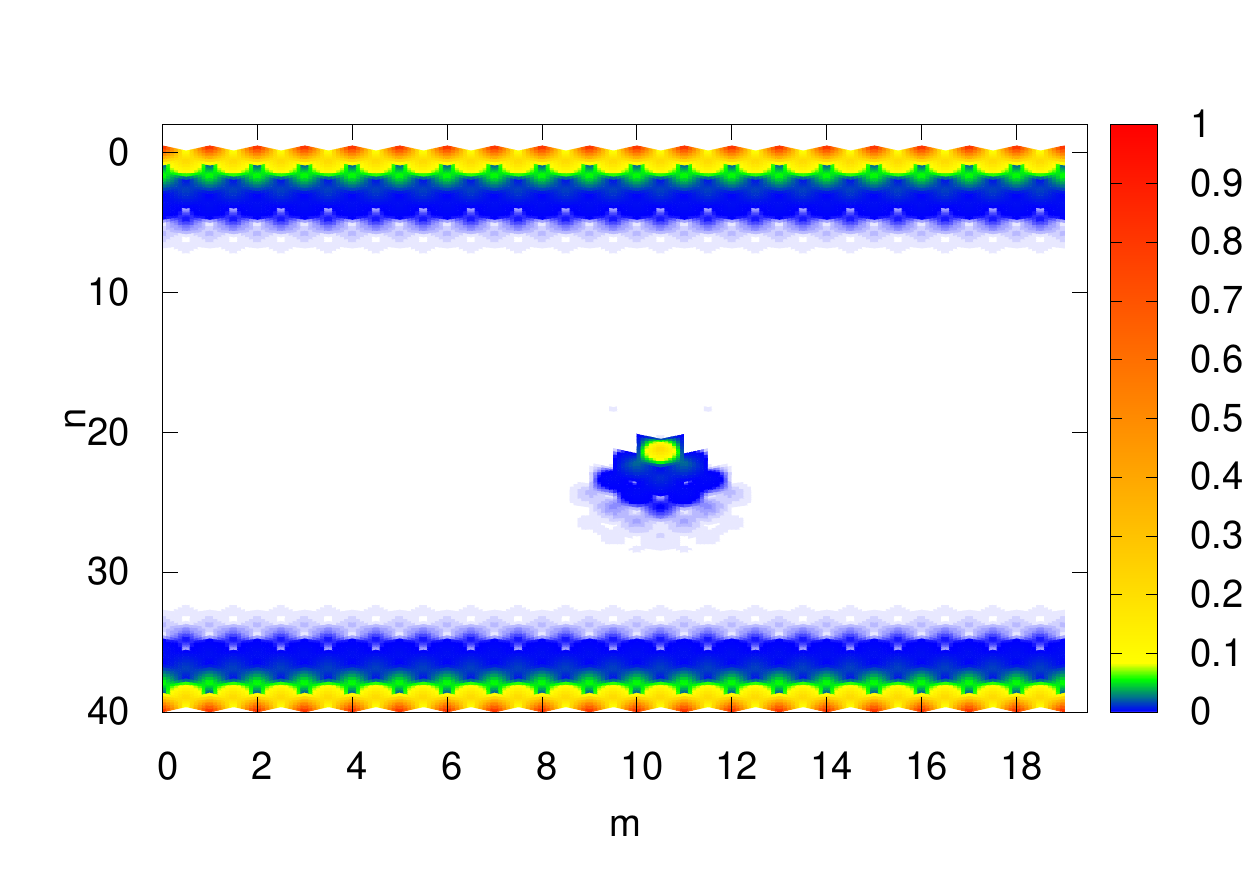}
\caption{(Color online). 
Contour plot of the amplitudes of a vacancy state which is located far away from the edges of a zPNR with $(m=20)$ armchair and $(n=40)$ zigzag chains. As a comparison the corresponding amplitudes of edge states is also plotted.} \label{Fig03}
\end{figure}
%%%%%%%%%%%%%%%%%%%%%%%%%%%%%%%%%%%%%%%%%%%%%%%%%%%%%%%%%%%%%%%%%%%%%%%%%%%%%%

However, before ending this section, let us consider a finite ribbon of phosphorene with zigzag boundary and locate a vacancy far away from the edges. 
Fig.~\ref{Fig03}, represents the corresponding LDOS of such a vacancy state which is obtained numerically in agreement with the analytical solution. We can compare our results with the LDOS of the edge states shown along the zigzag edges of the ribbon.
%As it is evident our analytical expression which is in excelent agreement with numerical results,  shows an anisotropic localization of vacancy wave function in real space in a zPNR which is very similar to the experimental observation of Ref.~\cite{Katsnelson2017} for bulk BP.

%%%%%%%%%%%%%%%%%%%%%%%%%%%%%%%%%%%%%%%%%%%%%%%%%%%%%%%%%%%%%%%%%%%%%%%%%%%%%%%
%%%%%%%%%%%%%%%%%%%%%%%%%%%%%%%%%%%%%%%%%%%%%%%%%%%%%%%%%%%%%%%%%%%%%%%%%%%%%%%
%%%%%%%%%%%%%%%%%%%%%%%%%%%%%%%%%%%%%%%%%%%%%%%%%%%%%%%%%%%%%%%%%%%%%%%%%%%%%%%
%%%%%%%%%%%%%%%%%%%%%%%%%%%%%%%%%%%%%%%%%%%%%%%%%%%%%%%%%%%%%%%%%%%%%%%%%%%%%%%
%%%%%%%%%%%%%%%%%%%%%%%%%%%%%%%%%%%%%%%%%%%%%%%%%%%%%%%%%%%%%%%%%%%%%%%%%%%%%%%
%%%%%%%%%%%%%%%%%%%%%%%%%%%%%%%%%%%%%%%%%%%%%%%%%%%%%%%%%%%%%%%%%%%%%%%%%%%%%%%%
\section{Fano resonances due to the coupling of a localized state with continuum states \label{Sec03}}

In order to study the phenomenon of Fano resonance, we can consider the impurity state as a quantum dot~(QD) coupled to a Q1D quantum system.   
The Hamiltonian describing the system is
\be
\label{EQ10}
\hat{H^{\prime}} = \hat{H_{Q1D}}+\hat{H_{QD}}+\hat{V}, 
\ee
where the first term is the Hamiltonian of a Q1D lattice with translational invariant which can be written using the plane wave basis $\vert k \rangle$ as
\be
\label{EQ11}
\hat{H_{Q1D}} =  \int_{-\frac{\pi}{a}}^{\frac{\pi}{a}} E(k) \vert k \rangle \langle k \vert dk,
\ee
in which $a$ is the lattice constant and $E(k)$ is the energy dispersion relation for electrons moving on the Q1D lattice in the absence of impurity.
The second term
\be
\label{EQ12}
\hat{H_{QD}} =  E_{0} \vert A \rangle \langle A \vert, 
\ee
describes a QD with energy $E_0$ and, finally, the last term describes the coupling between QD and Q1D as follows
\be
\label{EQ13}
\hat{V} = \int_{-\frac{\pi}{a}}^{\frac{\pi}{a}} \lbrace u(k) \vert k \rangle \langle A \vert + hc. \rbrace dk,
\ee
in which $u(k)$ is the coupling strength at momentum $k$. 
From now on, we set the lattice spacing to one, $a=1$.
%\begin{equation}
%\label{EQ9}
%\begin{split}
%\hat{H_{QW}} &=  \int_{-\frac{\pi}{a}}^{\frac{\pi}{a}} E(k) \vert k \rangle \langle k \vert dk, \\
%\hat{H_{QD}} &=  E_{0} \vert A \rangle \langle A \vert, \\
%\hat{V} &= \int_{-\frac{\pi}{a}}^{\frac{\pi}{a}} \lbrace u(k) \vert k \rangle \langle A \vert + H.c. \rbrace dk
%\end{split}
%\end{equation} 
%in which the first term is the momentum representation of kinetic energy and $a$ is tha lattice constant. 
%The second term describes the QD with energy $E_0$ and $u(k)$ is the tunneling coupling between QD and a Bloch state with momentum $k$. 

Since we are interested in the transmission coefficient $T(E)=|\mathcal{T}|^2$ (where $\mathcal{T}$ is the transmission amplitude), we can start with the Lippmann-Schwinger equation in the scattering theory
\begin{equation}
\label{EQ14}
\vert \psi_{out} \rangle = \vert \psi_{in} \rangle + \hat{G}_{0} \hat{T} \vert \psi_{in} \rangle,
\end{equation} 
where
\begin{equation}\label{EQ15}
%\begin{split}
\hat{T}(E) = \hat{V}\hat{G}_0(E) \hat{V} + \hat{V}\hat{G}_0(E) \hat{V}\hat{G}_0(E) \hat{V} + ..., %\\
%&= \hat{V} \big( 1 - \hat{G}_0(E) \hat{V} \big)^{-1}.
%\end{split}
\end{equation}
in which $\hat{G}_0(E)$ is the Green's operator in the absence of coupling between Q1D and QD. 
Since $\langle A\vert k\rangle=0$, it is formally possible to write
\begin{equation}
\label{EQ16}
\hat{G}_{0}(E) = \hat{G}_{0}^{Q1D}(E) + \hat{G}_{0}^{QD}(E),
\end{equation} 
in which 
\begin{equation}
\label{EQ17}
\hat{G}_{0}^{Q1D}(E) = \int_{-\pi}^{\pi} \frac{\vert k \rangle \langle k \vert}{E - E(k) + i0^{+}} dk,
\end{equation} 
and
\begin{equation}
\label{EQ18}
\hat{G}_{0}^{QD} = \frac{\vert A \rangle \langle A \vert}{E - E_{0} + i0^{+}},
\end{equation}
are the Green's functions of clean Q1D and QD respectively. 
With these expressions in hand, the first term of the series in Eq.~(\ref{EQ15}) can be evaluated as
\be
\label{EQ19}
\hat{V}\hat{G}_0(E) \hat{V} =(\vert A \rangle \langle A \vert) g+  \big(\int_{-\pi}^\pi(\vert k^\prime \rangle \langle k \vert) u(k)u(k^\prime) dkdk^\prime\big)g^\prime
\ee
where 
\begin{equation}
\label{EQ20}
g(E)= \int_{-\pi}^\pi \frac{\vert u{(k)} \vert ^{2}}{E - E{(k)} + i0^{+}} dk,
\end{equation} 
and
\begin{equation}
\label{EQ21}
g^\prime(E)= \frac{1}{E - E_{0} + i0^{+}}.
\end{equation} 
Doing the same analysis for the second term in Eq.~(\ref{EQ15}) and collecting these equations, we find,
\begin{equation}
\label{EQ22}
\hat{T}(E) = \frac{\hat{V} + \hat{V}^{\prime}}{1 - g(E) g^{\prime}(E)},
\end{equation} 
where we defined
\begin{equation}
\label{EQ23}
\hat{V}^{\prime} = g^{\prime} \int_{-\pi}^\pi u{(k)} u{(k^{\prime})} \vert k^{\prime} \rangle \langle k \vert dk \, dk^{\prime} + g (\vert A \rangle \langle A \vert).
\end{equation} 
According to Eq.~(\ref{EQ14}), the transmission amplitude at energy $E$ can be obtained as,
\begin{equation}
\label{EQ24}
\mathcal{T}(E) = 1 + \langle k_{0} \vert \hat{G}_{0} \hat{T} \vert k_{0} \rangle
\end{equation}
for an incident state $\vert k_0 \rangle $ such that  $E(k_0)=E$, hence,
\begin{equation}
\label{EQ25}
\mathcal{T}(E) = 1 + \frac{\frac{\vert u{(k_{0})} \vert ^{2}}{E-E_{0}} \int_{-\pi}^\pi \frac{1}{E-E{(k)}+i0^{+}}dk}{ 1 - \frac{1}{E-E_{0}} \int_{-\pi}^\pi \frac{\vert u{(k)} \vert ^{2}}{E-E{(k)}+i0^{+}}dk}.
\end{equation}

By having a quick look at the final expression for $\mathcal{T}(E)$, we can see that the coupling of an extra state (due to the presence of QD) with the continuum states with the same energy can induce an antiresonance in the transmission amplitude.  We will use this formalism in the next section to find this kind of resonant scattering due to the presence of vacancies in zPNRs.

\subsection{Antiresonance scattering of a 1D chain coupled to a QD}

\begin{figure}[t]
\centering
\includegraphics[scale=0.4]{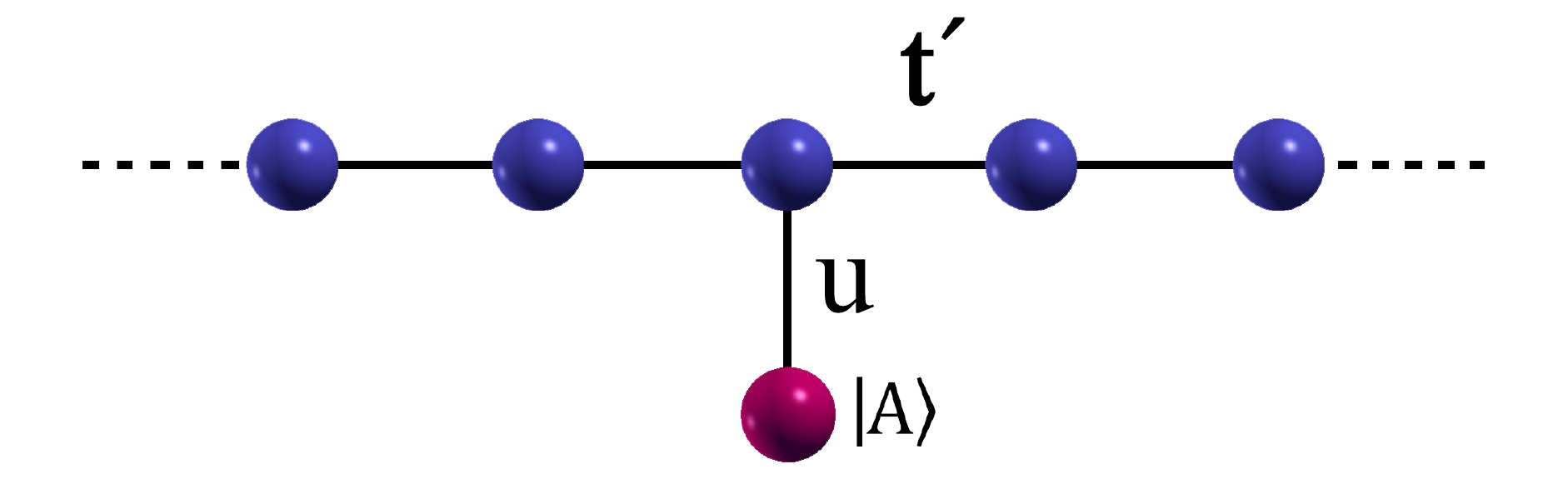}
\caption{(Color online). Schematic representation of a 1D chain with hopping amplitude $t'$ to the nearest neighbors.
An impurity state $\vert A\rangle$ is coupled with amplitude $u$ to the chain.} \label{Fig04}
\end{figure}

There is, however, at least one task remaining, namely, to show that the above formalism works fine.
In order to check our formalism, let us consider the trivial example of scattering of an electron traveling along a 1D chain by a QD (as shown in Fig.~\ref{Fig04}) which can be described by the Fano–Anderson tight-binding Hamiltonian~\cite{FlachRMP}. 
The energy dispersion of this model is given by
\begin{equation}
\label{EQ26}
E(k) = 2t' \, \cos{(k)} 
\end{equation}
where $t'$ is the hopping energy cost of an electron to the nearest neighbor sites.
We also consider a QD which is attached to the chain with on-site energy $E_0$ and coupling constant $u(k)=u$. 
%In this model, the hopping energy cost to the nearest neighbors is $t'$ and the on-site energy of QD is $E_0$ and coupling constant $u_0$ to the conduction states of the chain as shown in Fig.~\ref{Fig02}.    
Therefore, the resulting transmission is obtained by substituting these parameters into Eq.~(\ref{EQ25}) which yields the following expression 
%which can be used in  Eq.~(\ref{EQ22}) to reach the following expression for the transmission coefficient of the system
\begin{equation}
\label{EQ27}
\mathcal{T}(E) = 1 + \frac{\frac{u^{2}}{2it^{\prime} \sin{(k_0)}} \times \frac{1}{E-E_{0}}}{ 1 - \frac{u^{2}}{2it^{\prime} \sin{(k_0)}} \times \frac{1}{E-E_{0}}}
\end{equation} 
where $E_0=E(k_0)$.
Hence, we arrive at the exact solution~\cite{Orellana} for the Fano antiresonance in the transmission amplitude of this system.

%%%%%%%%%%%%%%%%%%%%%%%%%%%%%%%%%%%%%%%%%%%%%%%%%%%%%%%%%%%%%%%%%%%%%%%%%%%%%%%
%%%%%%%%%%%%%%%%%%%%%%%%%%%%%%%%%%%%%%%%%%%%%%%%%%%%%%%%%%%%%%%%%%%%%%%%%%%%%%

%Vacancies as resonant scatterers
%%%%%%%%%%%%%%%%%%%%%%%%%%%%%%%%%%%%%%%%%%%%%%%%%%%%%%%%%%%%%%%%%%%%%%%%%%%%%%%
%%%%%%%%%%%%%%%%%%%%%%%%%%%%%%%%%%%%%%%%%%%%%%%%%%%%%%%%%%%%%%%%%%%%%%%%%%%%%%%
%%%%%%%%%%%%%%%%%%%%%%%%%%%%%%%%%%%%%%%%%%%%%%%%%%%%%%%%%%%%%%%%%%%%%%%%%%%%%%%%
\section{Fano resonances in vacant zPNRs \label{Sec04}}
\begin{figure}[t]
\centering
\includegraphics[scale=0.8]{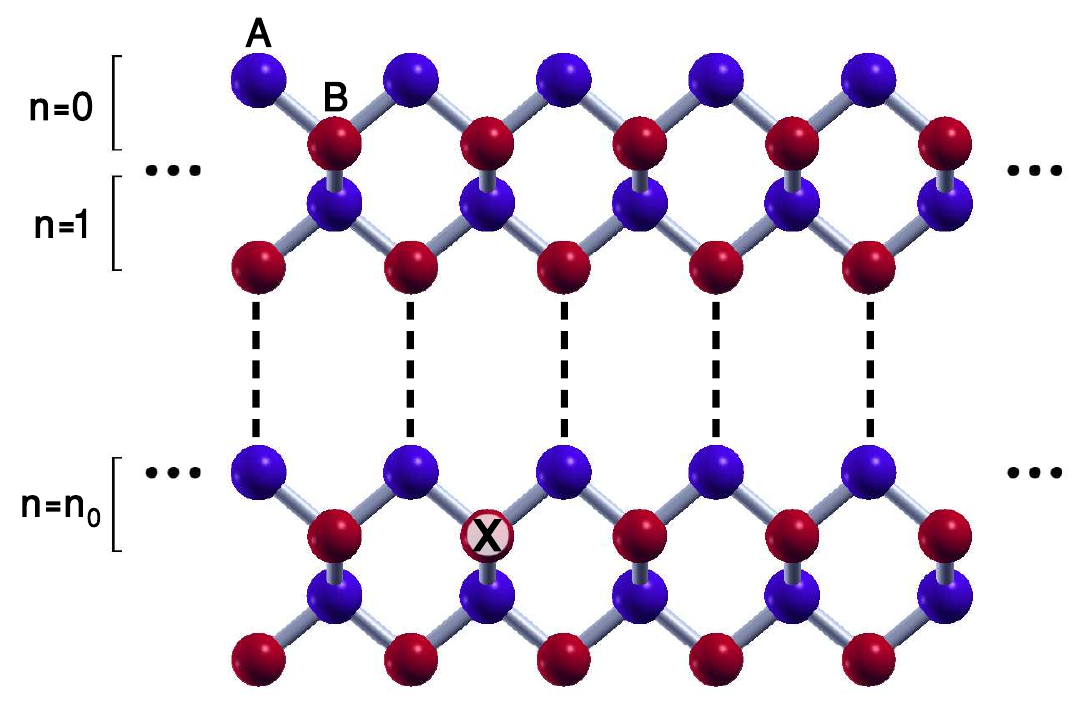}
\caption{(Color online). Schematic representation of a zPNR lattice structure with a single vacancy located on the $n_0$th zigzag chain with respect to the edge.} \label{Fig05}
\end{figure}

%\subsection{Fano resonance in vacant zPNRs}
In this section,  we use the above formalism to study the Fano resonance phenomenon which is induced by the presence of a single vacancy defect in ZPNRs.
In so doing, our approach is very similar to what was followed in the previous section for a 1D chain coupled to a QD.
Let us consider a single vacancy which is located at the $n_0$th zigzag chain with respect to the edge of ribbon as shown in
Fig.~\ref{Fig05}.

%%%%%%%%%%%%%%%%%%%%%%%%%%%%%%%%%%%%%%%%%%%%%%%%%%%%%%%%%%%%%%%%%%%%%%%%%%%%%%%
%%%%%%%%%%%%%%%%%%%%%%%%%%%%%%%%%%%%%%%%%%%%%%%%%%%%%%%%%%%%%%%%%%%%%%%%%%%%%%

As we already discussed, this vacancy introduces an additional impurity state whose wave function is $\vert\Psi_v^A\rangle$.
Now, the significant idea is that we would replace the role of Q1D and QD systems by the edge modes of the zPNR and vacancy state respectively.
Accordingly, we need to replace the required quantities $E(k)$, $E_0$ and $u(k)$ in Eq.~(\ref{EQ25}) with their counterparts of the vacant zPNR. This will be demonstrated shortly.

Let us start with first quantity $E(k)$ which is the electronic energy dispersion relation of the edge states in zPNR which is obtained in Ref.~\cite{Amini2018} as
\be
\label{EQ28}
E(k)=\langle \Psi^A(k)|\hat{H_1}|\Psi^A(k) \rangle =-2t'(1+\cos{(k)}). 
\ee
%where $\varepsilon_0=-4\frac{t_4t_1}{t_2}$ is an energy shift.
Before proceeding to the other quantities, we need to solve a technical problem.
It is important to note that according to Eqs.~(\ref{EQ4}) and (\ref{EQ5}), the states $\vert \Psi^A(k)\rangle$ and  $\vert\Psi_v^A\rangle$ are not orthogonal, namely $\langle \Psi^A(k)\vert\Psi_v^A\rangle\neq0$. 
Thus, we need to define an orthogonal state $\vert{\Psi^\prime}_v^A\rangle$ as
\begin{equation}
\label{EQ29}
\vert{\Psi^\prime}_v^A\rangle = c^{\prime} \left(\vert \Psi_v^A\rangle - \int_{-\pi}^{\pi} \langle \Psi_v^A | \Psi^A(k) \rangle | \Psi^A(k) \rangle dk \right)
\end{equation} 
such that the required condition $\langle \Psi^A(k)\vert{\Psi^\prime}_v^A\rangle = 0$ is satisfied.
Here, 
\be
\label{EQ30}
(c^\prime)^{-2} =1- c^{-2} \int_{-\pi}^{\pi} \alpha^4(k) \gamma^{-2}(k) dk 
\ee
is a constant that can be used to normalize the wave function.
It is now possible to obtain the energy of this bound state which resembles the quantity $E_0$.
Due to the presence of the term $\hat{H_1}$  the corresponding bound-state energy of $\vert{\Psi^\prime}_v^A\rangle$ is changed from zero and can be obtained perturbatively as
%the corresponding energy of $\vert{\Psi^\prime}_v^A\rangle$ due to the presence of $\hat{H_1}$ is changed and can be obtained 
\begin{equation}
\label{EQ31}
E^{\prime}_0 = \langle {\Psi^\prime}_v^A | \hat{H_{1}} |{\Psi^\prime}_v^A \rangle =\left( E_{00} - E_{01} - E_{10}^* + E_{11} \right),
\end{equation} 
where $E_{ij}$s are given by
\be
\label{EQ32}
E_{01}=c^2\int_{-\pi}^{\pi} 2t^\prime\gamma^{-2}(k)\cos^2{(k/2)} \left[ \alpha^{2n_0-2}(k)+\alpha^{2n_0}(k) \right], 
\ee
and 
\be
\label{EQ33}
E_{11}=c^2\int_{-\pi}^{\pi} 4t^\prime\gamma^{-2}(k)\cos^2{(k/2)}  \alpha^{2n_0}(k). 
\ee
$E_{00}$ is given by Eq.~(\ref{EQ9}) and $E_{10}^*=E_{01}$.
%There is an important point about the values of bound state energy $E^\prime_0$ that should be noticed when the vacancy position changes.
It is now important, however, to note that if the bound state energy $E^\prime_0$ lies within the energy interval of the edge modes, namely $-2t^\prime<E^\prime_0<0$, then it is expected to observe the Fano antiresonance for the edge modes. 
For instance,  
\begin{subequations}
	\label{EQ34}
	\begin{equation}
E^{\prime}_0 = 0.00 \,\,\,\, for \,\,\,\, n_0=0
	\end{equation}
	\begin{equation}
E^{\prime}_0 = -0.11 \,\,\,\, for \,\,\,\, n_0=1
	\end{equation}
	\begin{equation}
E^{\prime}_0 = -0.14 \,\,\,\, for \,\,\,\, n_0=2
	\end{equation}
	\begin{equation}
E^{\prime}_0 = -0.15 \,\,\,\, for \,\,\,\, n_0=3.
	\end{equation}
\end{subequations}
which shows that for $n_0=0$ one can not observe the Fano antiresonance because it's corresponding bonding energy lies outside the bandwidth.

We are now in a position to calculate the only remaining quantity $u(k)$ which is the coupling parameter between the vacancy state and edge states, namely
%It can be expressed in terms of the new vacancy state as 
\begin{equation}
\label{EQ35}
u(k) = \langle k | \hat{H_{1}} | {\Psi^\prime}_v^A \rangle.
\end{equation} 
This results in
\begin{equation}
\label{EQ36}
u(k) = {c^\prime}^2 \gamma^{-2}({k})\frac{t_2t_4}{t_1} \left[\alpha^{2n_0}({k}) -2 \alpha^{2n_0+2}({k}) + \alpha^{2n_0+4}({k})  \right].
\end{equation} 
Collecting these results and inserting into Eq.~(\ref{EQ25}), we can obtain the transmission amplitude through the edge states of a zPNR in presence of a single vacancy which is located on $n_0$th zigzag chain.
Thus,
\begin{equation}
\label{EQ37}
\mathcal{T}{(E)} = 1 + \frac{\frac{u^{2}{(k_{0})}}{2it^{\prime} \sin(k_{0})} \times \frac{1}{E-E^{\prime}_{0}}}{1 - \frac{1}{E-E^{\prime}_{0}} \times \int_{-\pi}^{\pi} \frac{u^{2}{(k)}}{E-2t^{\prime} \cos^{2}(k/2) + i0^{+}}dk}
\end{equation} 
where $E(k_0)=E$ and the integration can be carried out analytically using contour integration~\cite{Amini2018,Amini2018-2}.
%%%%%%%%%%%%%%%%%%%%%%%%%%%%%%%%%%%%%%%%%%%%%%%%%%%%%%%%%%%%%%%%%%%%%%%%%%%%%
%%%%%%%%%%%%%%%%%%%%%%%%%%%%%%%%%%%%%%%%%%%%%%%%%%%%%%%%%%%%%%%%%%%%%%%%%%%%%
%%%%%%%%%%%%%%%%%%%%%%%%%%%%%%%%%%%%%%%%%%%%%%%%%%%%%%%%%%%%%%%%%%%%%%%%%%%%%
\section{Results and discussion \label{Sec05}}

%%%%%%%%%%%%%%%%%%%%%%%%%%%%%%%%%%%%%%%%%%%%%%%%%%%%%%%%%%%%%%%%%%%%%%%%%%%%%%%
\begin{figure}[t]
\centering
\includegraphics[scale=0.7]{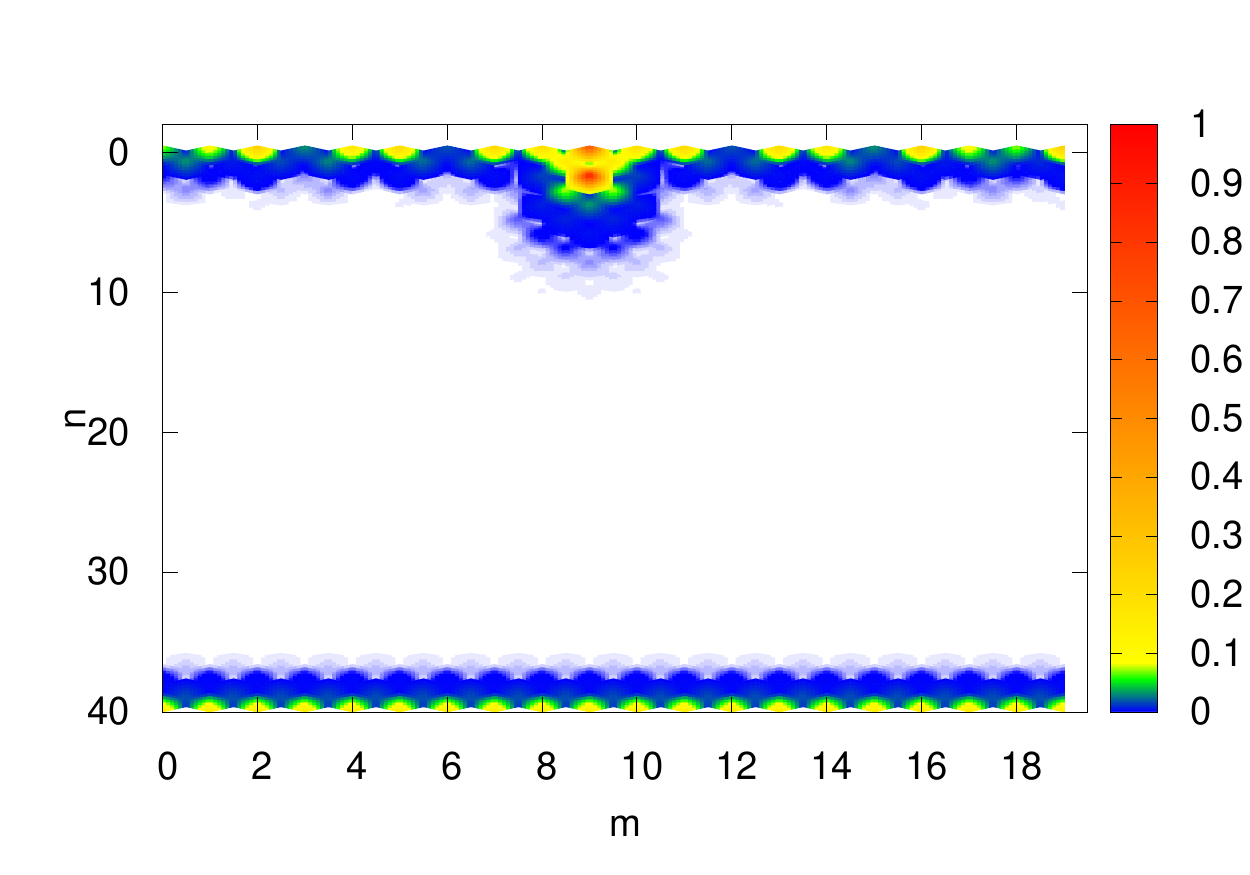}
\caption{(Color online). Snapshot of the LDOS of edge states and localized state around a single vacancy which is located on the  $n_0=1$ zigzag chain of a zPNR with 1600 atoms. } \label{Fig06}
\end{figure}
%%%%%%%%%%%%%%%%%%%%%%%%%%%%%%%%%%%%%%%%%%%%%%%%%%%%%%%%%%%%%%%%%%%%%%%%%%%%%%

%%%%%%%%%%%%%%%%%%%%%%%%%%%%%%%%%%%%%%%%%%%%%%%%%%%%%%%%%%%%%%%%%%%%%%%%%%%%%%%
\begin{figure}[t]
\centering
\includegraphics[scale=0.7]{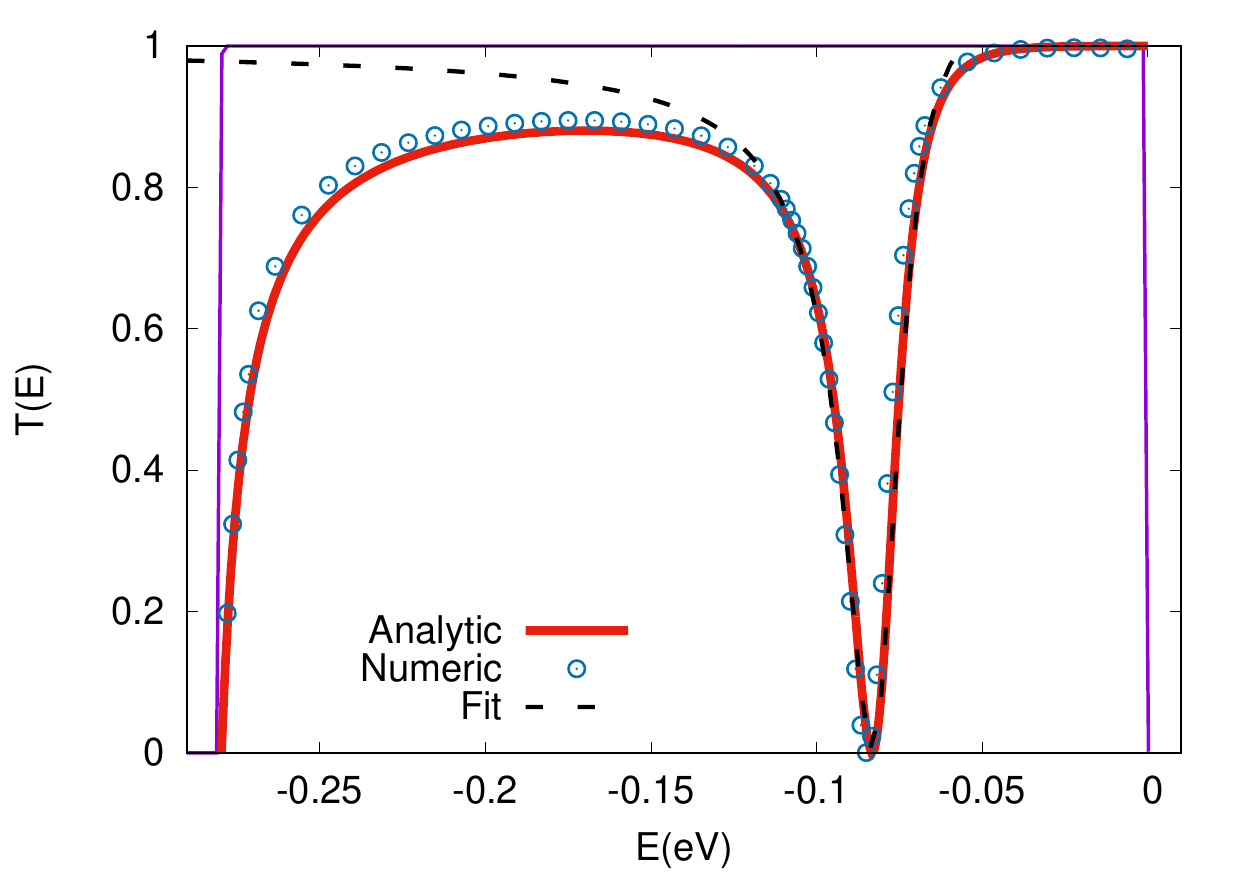}
\caption{(Color online).
Transmission coefficient of an electron traveling along the zigzag edge of a zPNR in presence of a single vacancy which is located on site $(m=0,n_0=1,B)$ as a function of incident electron energy.
The solid lines corresponds to the analytical results, the black dahsed lines shows the standard Fano line shape fitted to the analytical expression close to the resonance energy, and the discrete circles corresponds to the numerical results obtained using the recursive Green's function method.} \label{Fig07}
\end{figure}
%%%%%%%%%%%%%%%%%%%%%%%%%%%%%%%%%%%%%%%%%%%%%%%%%%%%%%%%%%%%%%%%%%%%%%%%%%%%%%

In this section we consider a zPNR in presence of a single vacancy which is located at site $(m=0,n_0,B)$.
As we already discussed, only when $n_0\geq1$ the Fano antiresonance takes place. Therefore, 
for simplicity, we first assume $n_0=1$ which is the first place in which the vacancy can induce antiresonance.
In Fig.~\ref{Fig06} we show the corresponding LDOS of the edge states and localized vacancy state obtained numerically. 
In order to obtain the transmission amplitude of such a system, we can insert Eqs.~(\ref{EQ31}) and (\ref{EQ36}) into Eq.~(\ref{EQ37}) which results in 
%\begin{equation}
%\label{EQ38}
%$\int_{-\pi}^{\pi} \frac{u^{2}(k)}{E-2t^{\prime} \cos^{2}(\frac{k}{2}) + i0^{+}}dk$
%\end{equation} 
%The result is,
\begin{equation}
\label{EQ38}
\mathcal{T}{(E)} = 1 + \frac{\frac{|u{(k_{0})}|^{2}}{2it^{\prime} \sin(k_{0})} \times \frac{1}{E-E^{\prime}_{0}}}{1 - \frac{1}{E-E^{\prime}_{0}} \times \frac{\beta{(k_0)}}{2it^{\prime} \sin(k)}}
\end{equation} 
in which $E^\prime_0$ and $u(k)$ should be evaluated for $n_0=1$ and $\beta{(k_0)}$ is rather a complicated expression which we do not present here. Instead, we present a graphical representation of the transmission coefficient $T(E)$ using the transmission amplitude in Eq.~(\ref{EQ38}) as a function of energy $E$ in Fig.~\ref{Fig07}.
It is clearly shown in Fig.~\ref{Fig07} that there exists a Fano antiresonance around resonance energy $E_r(n_0=1)\approx-0.08$~(and close to it's corresponding bound state energy $E^\prime_0$) for which the transmission through the edge channel vanishes. This is due to the destructive interference of the traveling wave which can either bypass the vacancy state or populate it and return back to the continuum.
Let us now check the deviation of represented expression from the standard line shape of the Fano resonances~\cite{FlachRMP} of the form 
\be
\label{EQ39}
T(\varepsilon)=\frac{(\varepsilon+q)^2}{\varepsilon^2+1},
\ee
where $\varepsilon=(E-E_r)/(\Gamma/2)$
and $\Gamma$, $E_r$, and $q$ are the line width, peak position, and asymmetry parameter respectively.
Fig.~\ref{Fig07} shows the fitted curve~(dashed lines) with this Fano formula which shows an excellent agreement to a
Fano line shape with asymmetry parameter $q=0.18$.
This difference with the case of an exactly 1D chain which shows a symmetric transmission profile is reasonable since the zPNR is a Q1D system and the edge states have non-zero amplitudes within the localization radius near the edge.
In order to check our analytical expression, we also use the numerical Landauer approach that works on
the basis of the recursive Green’s function technique and we have used it in Ref.~\cite{Amini2018}.
Fig.~\ref{Fig07} (discrete circles) shows the resulting transmission which is obtained from the numerical calculation and is in perfect agreement with the analytical solution.

\begin{figure}[t!]
\centering
\includegraphics[scale=0.7]{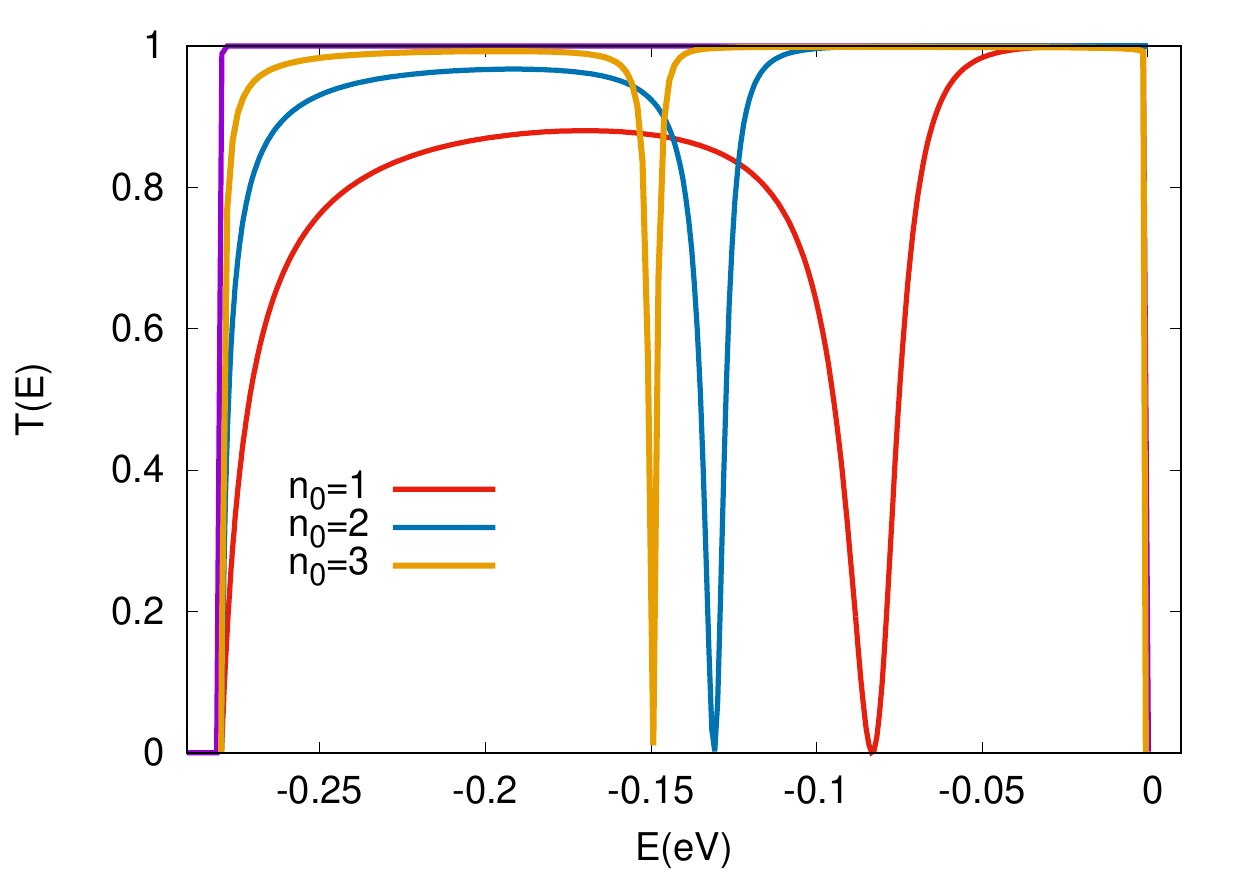}
\caption{(Color online).
The dependence of resonant peak shape and position on the  vacancy distance from the edge of zPNR plotted  with different colors: 
red $n_0=1$, blue $n_0=2$, and orange $n_0=3$.} \label{Fig08}
\end{figure}

Next, we study the influence of varying the vacancy position on the Fano lineshape properties. To this end, we change the vacancy position to different distances from the edge, namely increasing $n_0$ to $n_0=2,3$.  
Fig.~\ref{Fig08} presents the transmission probability versus the incident electron energy for various vacancy positions.
It is clear that displacing the vacancy away from the edge of the zPNR causes several important differences: 
(i) a shift in the resonance energy toward the center of the edge band; 
(ii) suppression of the asymmetry parameter $q$, and (iii) decreasing the line width of the peak, $\Gamma$.
The reason is that by increasing the distance of vacancy from the edge, the coupling parameter $u(k)$ decreases due to the suppression of wave functions overlap and since the resonance width is directly proportional to the coupling strength, therefore $\Gamma$ decreases. Furthermore, far away from the edge, one can consider the edge states as an exactly 1D chain, therefore, the transmission profile should be symmetric and locate at the band center.
Before ending this discussion, we should emphasize that our results are completely consistent with the results of Ref.~\cite{Peeters2018}.

%%%%%%%%%%%%%%%%%%%%%%%%%%%%%%%%%%%%%%%%%%%%%%%%%%%%%%%%%%%%%%%%%%%%%%%%%%%%%%%
%%%%%%%%%%%%%%%%%%%%%%%%%%%%%%%%%%%%%%%%%%%%%%%%%%%%%%%%%%%%%%%%%%%%%%%%%%%%%%

\begin{figure}[t]
	\centering
	\includegraphics[scale=0.7]{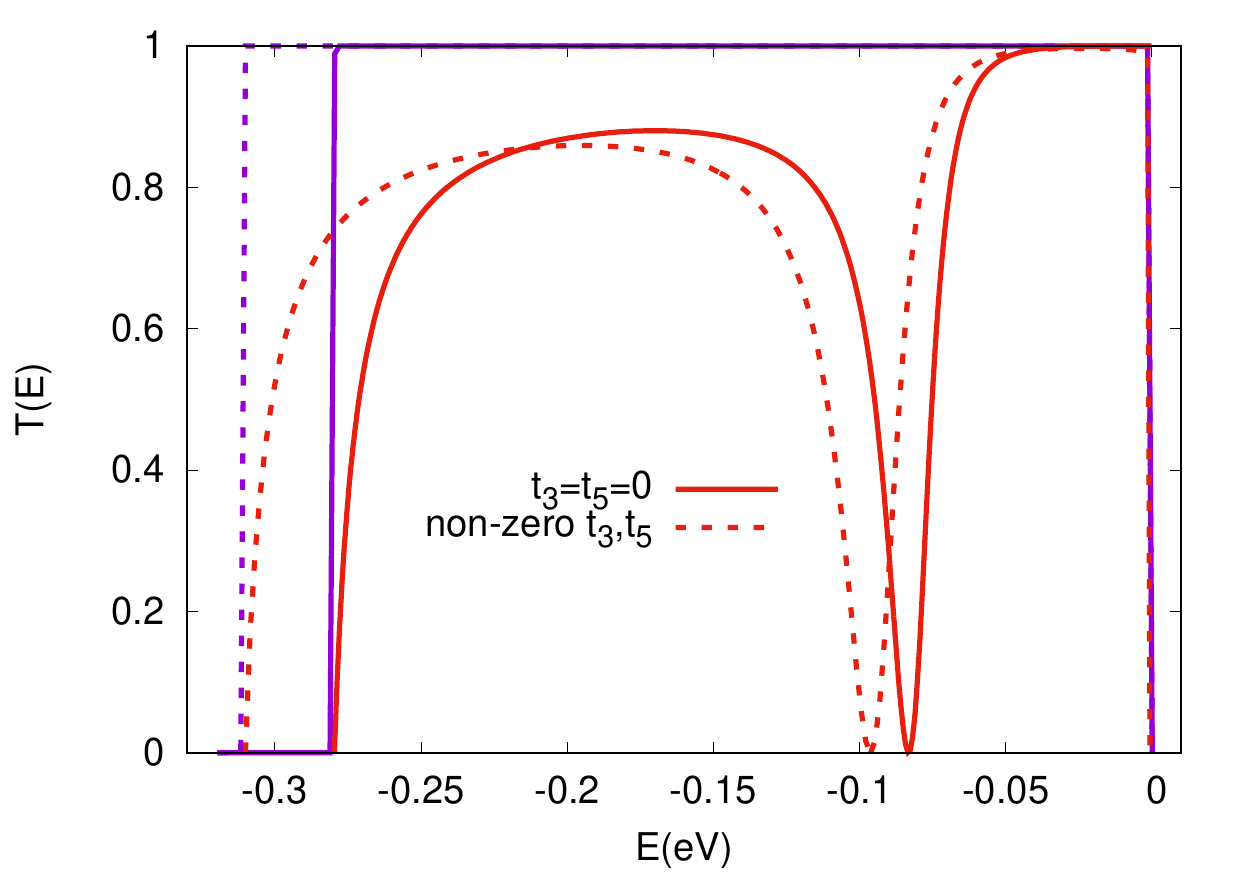}
	\caption{Color online).
		The effect of none-zero hopping integrals $t_3$ and $t_5$ on the resonant peak positions induced by a single vacancy which is located at $n_0=1$ zigzag chain. The dashed lines obtained in the presence of $t_3$ and $t_5$ and compared with their counterparts with solid lines when $t_3 = t_5 = 0$.} \label{Fig09}
\end{figure}

Finally, it is worth pointing out that the presence of hopping parameters $t_3$ and $t_5$ does not change the overall picture of the transmission coefficient in general. To clarify this point, we present a plot of transmission coefficient versus energy for both cases, namely the case in which $t_3=t_5=0$ and the case with non-zero $t_3$ and $t_5$, in Fig.~\ref{Fig09}.
As presented in Fig.~\ref{Fig09}, the resonant peaks of transmission coefficient in presence of $t_3$ and $t_5$ (which is shown with dashed lines) takes place with a small energy shift. This is reasonable because the presence of $t_3$ and $t_5$ increases the energy bandwidth of the edge band.

%%%%%%%%%%%%%%%%%%%%%%%%%%%%%%%%%%%%%%%%%%%%%%%%%%%%%%%%%%%%%%%%%%%%%%%
%%%%%%%%%%%%%%%%%%%%%%%%%%%%%%%%%%%%%%%%%%%%%%%%%%%%%%%%%%%%%%%%%%%%%%%
\section{Summary \label{Sec06}}
 In this paper, we have demonstrated the existence of a new type of impurity state associated with vacancies in monolayer phosphorene which exhibits a strongly anisotropic localization profile in real space.
 We found that the localized state induced by vacancies can overlap with the wave functions of the edge states in a confined ribbon of phosphorene which results in the formation of dips in the electronic transmission through the edge modes.  The latter takes place when the energy of the vacancy bound state lies inside the quasi flat band composed of the edge modes.
 We have derived analytically and validated numerically the theory of such antiresonances in zPNRs with a vacancy using the T-matrix Lippmann-Schwinger approach. 
 Modeling the continuum states of the edges of zPNRs as a Q1D system and the vacancy state as a QD, we found that it is exactly a Fano phenomenon due to the coupling of vacancy state to the continuum of the edge modes.  
 We have tried to shed light on various effects of displacing the vacancy
 on such Fano antiresonances that occur in a zPNR with single vacancies.

%%%%%%%%%%%%%%%%%%%%%%%%%%%%%%%%%%%%%%%%%%%%%%%%%%%%%%%%%%%%%%%%%%%%%%%%%%%%%%%
%%%%%%%%%%%%%%%%%%%%%%%%%%%%%%%%%%%%%%%%%%%%%%%%%%%%%%%%%%%%%%%%%%%%%%%%%%%%%%

%%%%%%%%%%%%%%%%%%%%%%%%%%%%%%%%%%%%%%%%%%%%%%%%%%%%%%%%%%%%%%%%%%%%%%%%
\begin{acknowledgments}
We gratefully acknowledge discussion with F. M. Peeters and L. L. Li.
MA acknowledges the support of the Abdus Salam (ICTP) associateship program.
MSh would also like to acknowledge the office of graduate
studies at the University of Isfahan for their support and research
facilities.
\end{acknowledgments}
%%%%%%%%%%%%%%%%%%%%%%%%%%%%%%%%%%%%%%%%%%%%%%%%%%%%%%%%%%%%%%%%%%%%%%%%
%%%%%%%%%%%%%%%%%%%%%%%%%%%%%%%%%%%%%%%%%%%%%%%%%%%%%%%%%%%%%%%%%%%%%%%%%%%%%%%

%%%%%%%%%%%%%%%%%%%%%%%%%%%%%%%%%%%%%%%%%%%%%%%%%%%%%%%%%%%%%%%%%%%%%%%%%%%%%%%


\begin{thebibliography}{}

\bibitem{Li2014} L. Li, Y. Yu, G. J. Ye, Q. Ge, X. Ou, H. Wu, D. Feng, X. H.
Chen, and Y. Zhang, Nat. Nanotechnol.~\textbf{9}, 372 (2014).

\bibitem{Li2016} L. Li, F. Yang, G. J. Ye, Z. Zhang, Z. Zhu, W. Lou, L. Li, X. Zhou, K. Watanabe, T. Taniguchi, K. Chang, Y. Wang, X. H. Chen, and Y. Zhang, Nat. Nanotechnol. ~\textbf{11}, 593 (2016).

\bibitem{Liu2014} H. Liu, A. T. Neal, Z. Zhu, Z. Luo, X. Xu, D. Tomanek, and P. D.
Ye, ACS Nano~\textbf{8}, 4033 (2014).

\bibitem{Xia2014} F. Xia, H. Wang, and Y. Jia, Nat. Commun.~\textbf{5}, 4458 (2014).

\bibitem{Neto-2014} S. P. Koenig, R. A. Doganov, H. Schmidt, A. H. Castro Neto,
and B. Zyilmaz, Appl. Phys. Lett.~\textbf{104}, 103106 (2014).

\bibitem{Ling2015} X. Ling, H. Wang, S. Huang, F. Xia, and M. S. Dresselhaus, Proc. Natl. Acad. Sci. (USA) \textbf{112}, 4523 (2015).

\bibitem{Carvalho2016} Alexandra Carvalho, Min Wang, Xi Zhu, Aleksandr S. Rodin, Haibin Su and Antonio H. Castro Neto, Nat. Rev. Mater. \textbf{1}, 16061 (2016).


\bibitem{Qiao2014} J. Qiao, X. Kong, Z. X. Hu, F. Yang, and W. Ji, Nat. Commun.
\textbf{5}, 4475 (2014).



\bibitem{Neto2014} A. S. Rodin, A. Carvalho, and A. H.  Castro Neto, Phys. Rev. Lett.
\textbf{112}, 176801 (2014).



\bibitem{Peeters2014} D. Cakir, H. Sahin, and F. M. Peeters, Phys. Rev. B \textbf{90}, 205421
(2014).

%%%%%10
\bibitem{Guinea2014} T. Low, R. Roldan, H. Wang, F. Xia, P. Avouris, L. M. Moreno,
and F. Guinea, Phys. Rev. Lett. \textbf{113}, 106802 (2014).

\bibitem{Yang2014} R. Fei and L. Yang, Nano Lett. \textbf{14}, 2884 (2014).

\bibitem{Katsnelson2015} S. Yuan, A. N. Rudenko, and M. I. Katsnelson, Phys. Rev. B \textbf{91},
115436 (2015).

%\bibitem{Asgari2015} M. Elahi, K. Khaliji, S.M. Tabatabaei, M. Pourfath and R. Asgari, Phys. Rev. B \textbf{91}, 115412 (2015).

%\bibitem{Feng2015} Qingyun Wu, Lei Shen, Ming Yang, Yongqing Cai, Zhigao  Huang,  and  Yuan  Ping  Feng,  Phys.  Rev.  B \textbf{92}, 035436 (2015).

%\bibitem{Maity2016} Ajanta Maity, Akansha Singh, Prasenji Sen, Aniruddha Kibey,Anjali  Kshirsagar,  and  Dilip  G.  Kanhere,  Phys. Rev. B \textbf{94}, 075422 (2016).

%\bibitem{Fang2017} Fang  Xie,  Zhi-Qiang  Fan,  Xiao-Jiao  Zhang,  Jian-Ping Liu, Hai-Yan Wang,Kun Liu, Ji-Hai Yu, Meng-Qiu Long, Organic Electronics \textbf{42}, 21 (2017).

\bibitem{Ezawa2014} M. Ezawa, New Journal of Physics \textbf{16}, 115004 (2014).

\bibitem{Peeters-skew} M. M. Grujic, M. Ezawa, M. Z. Tadic, and F. M. Peeters,
Phys. Rev. B \textbf{93}, 245413 (2016).

\bibitem{Asgari2018} Z. Nourbakhsh and R. Asgari, arXiv:1803.00751(2018).

\bibitem{Amini2018} M. Amini and M. Soltani, arXiv:1810.03042 (2018).

\bibitem{Hu2015} W. Hu and J. Yang, J. Phys. Chem. C \textbf{119}, 20474 (2015).

\bibitem{Wang2015} V. Wang, Y. Kawazoe, and W. T. Geng, Phys. Rev. B \textbf{91}, 045433 (2015).

\bibitem{Katsnelson2017} B. Kiraly, N. Hauptmann, A. N. Rudenko, M. I. Katsnelson, and A. A. Khajetoorians, Nano Lett. \textbf{17}, 3607 (2017).

%%%%%%%%20


\bibitem{Peeters2018} L. L. Li, and F. M. Peeters, Phys. Rev. B \textbf{97}, 075414 (2018).

\bibitem{Fano} U. Fano Phys. Rev. \textbf{124}, 1866 (1961).

\bibitem{Rudenko2014} A. N. Rudenko, M. I. Katsnelson, Phys. Rev. B \textbf{89}, 201408 (2014).


\bibitem{Pereira2008} V. M. Pereira, J. M. B. Lopes dos Santos, and A. H. Castro Neto, Phys. Rev. B \textbf{77}, 115109 (2008).

\bibitem{Pereira2006} V. M. Pereira, F. Guinea, J. M. B. Lopes dos Santos, N. M. R. Peres, and A. H. Castro Neto, Phys. Rev. Lett. \textbf{96}, 036801 (2006).

\bibitem{Vozmediano} E. V. Castro, M. P. López-Sancho, and M. A. H. Vozmediano, Phys. Rev. Lett. \textbf{104}, 036802 (2010).

%\bibitem{Dutreix2013} C. Dutreix,  L. Bilteanu,  A. Jagannathan,  and C. Bena 87, Phys. Rev. B \textbf{87}, 245413 (2013).

\bibitem{Ezawa2018} M. Ezawa, Phys. Rev. B \textbf{98}, 045125 (2018).

\bibitem{FlachRMP} A. E. Miroshnichenko, S. Flach, and  Y. S. Kivshar Rev. Mod. Phys. \textbf{82} 2257 (2010).

\bibitem{Orellana} P. A. Orellana, F. Domínguez-Adame, I. Gómez, and M. L. Ladrón de Guevara, Phys. Rev. B \textbf{67}, 085321 (2003).

\bibitem{Amini2018-2} M. Amini, M. Soltani, E. Ghanbari-Adivi, and M. Sharbafiun arXiv:1810.03486 (2018).

%\bibitem{Ghanbari2013}     E. Ghanbari-Adivi,  M. Soltani, and H. Ebtekarnasab, Eur. Phys. J. D \textbf{67}, 118 (2013)

%\bibitem{Ghanbari2015}     E. Ghanbari-Adivi, M. Soltani, and M. N. Sheikhali, Eur. Phys. J. D \textbf{69}, 172 (2015)

%\bibitem{Ghanbari2016}     E. Ghanbari-Adivi, M. Soltani, and M. Sheikhali, Quantum Inf. Process. \textbf{15}, 2377 (2016)

%\bibitem{Economou1990} E. N. Economou, \emph{Green's function in
%quantum physics}, Springer Verlag, Berlin (2006)


\end{thebibliography}
\end{document}